\documentclass[runningheads]{llncs}
\usepackage{algorithm}
\usepackage[noend]{algpseudocode}
\usepackage{graphicx}
\usepackage{amssymb}
\usepackage{amsmath}
\usepackage{stmaryrd}
\usepackage{MnSymbol}
\usepackage{fancyhdr}
\usepackage{xcolor}
\usepackage{soul}

\makeatletter
\def\algbackskip{\hskip-\ALG@thistlm}
\makeatother
\newcommand{\UsedPeriods}{\textit{UsedPeriods}}
\newcommand{\AccumulatedPeriods}{\textit{AccumulatedPeriods}}
\newcommand{\newperiods}{\textit{NewPeriods}}
\newcommand{\eASTD}{\textsf{eASTD}}
\newcommand{\cASTD}{\textsf{cASTD}}

\newcommand{\mygraphic}[1]{\includegraphics[height=#1]{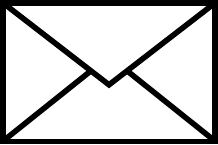}}
\newcommand{\myenv}{(\raisebox{0pt}{\mygraphic{.6em}})}
\newcommand{\myauthor}[1]{#1~\myenv}
\begin{document}
\title{Development of monitoring systems for anomaly detection using ASTD specifications}

%
%
\author{\myauthor{El Jabri Chaymae\inst{1}} \and
Frappier Marc\inst{1}\and
Ecarot Thibaud\inst{1} \and
Tardif Pierre-Martin \inst{2}}
\authorrunning{C. El Jabri et al.}
%
\titlerunning{Development of monitoring systems using ASTD spec}
\institute{
Computer Science Department at Université de Sherbrooke, GRIF, Québec, Canada\\
\email{\{chaymae.el.jabri,marc.frappier,thibaud.ecarot\}@usherbooke.ca}
\and Management School at Université de Sherbrooke, Québec, Canada\\
\email{pierre-martin.tardif@usherbrooke.ca}
}
\maketitle              
\begin{abstract}

Anomaly-based intrusion detection systems are essential defenses against cybersecurity threats because they can identify anomalies in current activities. However, these systems have difficulties providing entity processing independence through a programming language. In addition, a degradation of the detection process is caused by the complexity of scheduling the training and detection processes, which are required to keep the anomaly detection system continuously updated. This paper shows how to use the algebraic state-transition diagram (ASTD) language to develop flexible anomaly detection systems. This paper provides a model for detecting point anomalies using the unsupervised non-parametric technique Kernel Density Estimation to estimate the probability density of event occurrence. The proposed model caters for both the training and the detection phase continuously. The ASTD language streamlines the modeling of detection systems thanks to its process algebraic operators that provide a solution to overcome these challenges. By delegating the combination of anomaly-based detection processes to the ASTD language, the effort and complexity are reduced during detection models development. Finally, using a qualitative evaluation, this study demonstrates that the algebraic operators in the ASTD specification language overcome these challenges.

\end{abstract}
\section{Introduction}

Critical systems and sensitive infrastructure are increasingly subject to an intensification of cyberattacks, the complexity of which increases throughout multiple offensives. To adequately counter the risks that are not always identified and known, these systems must have a defense with the main characteristic of quickly and effectively detecting a threat or abnormal behavior. These threats, which are often composed of a variety of combined tactics and techniques that adversaries may employ to achieve their objectives, are increasingly challenging to detect due to their inherent heterogeneity and complexity~\cite{Thakkar2021}. 

Detecting these threats is challenging because of detecting heterogeneous attacks with various variants, the need to quickly obtain a representative and up-to-date dataset for the training phase, and the management of internal processes and alarm handling of an intrusion detection system (IDS). Indeed, IDSs must detect a wide range of attacks whose nature can vary within a given system substrate~\cite{Kasinathan2013DenialofServiceDI,RAZA20132661}. In addition, the management of the processes running within the detection systems is of great complexity, mainly due to the number of entities present in the ever-growing infrastructures, whose topology is constantly changing and which are deployed on a large scale~\cite{sanchez2011smartsantander}. It is also about getting representative datasets of these changes faster and processing a massive amount of generated alerts to reduce the number of false positives, or irrelevant alerts~\cite{khakurel_bhagat_2019,Thakkar2020ARO}. These various challenges have highlighted the difficulty of adapting IDSs to changes. These difficulties impact IDS based on dynamic signatures and anomaly detection.

To answer these difficulties of adaptation, particular works of~\cite{8369054,Zhang2019MultilayerDC} have examined several approaches, such as the combination of different learning techniques or the use of a better classification using labeling before the training phase. However, specific challenges persist with these works, particularly the lack of flexibility due to the process scheduling during the training and detection phases executed in parallel~\cite{9682357} and the entity profiles independence to be monitored~\cite{neucom}. More specifically, process scheduling is a persistent issue during IDS' development. Indeed, programming languages do not have predefined formal operations ensuring the interaction between the multiple processes, complicating continuous improvement and reducing reusability. The next challenge is the unique treatment of the entities of a system to be monitored. Indeed, the processing is unique for all the characteristics of the entities. It is impossible to differentiate the entities because the training and detection model parameters are specified a priori. A final issue is the interruption of detection when renewing training data because the feedback loop is not continuous or automated.

In order to answer appropriately to these functional issues, it was hypothesized that the use of the algebraic state transition diagrams (ASTDs) formalization language~\cite{astdarticle} would make it possible to meet effectively to these challenges. ASTD is an executable, modular and graphical notation that allows for the composition of hierarchical state machines using process algebra operators such as flow, sequence, quantified interleaving, and parallel synchronization~\cite{tidjon2020intrusion}. Indeed, using algebra operators specific to the IDS' development coming from this language should improve the reliability and flexibility of these systems. The research work presented in this article aims to formalize the development of intrusion detection systems and to achieve three objectives: 
\begin{itemize}
    \item Separate the coordination of the processes from the actions constituting the model; 
    \item Provide independent processing of each entity that constitutes the system to be monitored;
    \item Ensure continuous processing of events between the training and detection phase. 
\end{itemize}

This paper is structured as follows. Section~\ref{related-work} first explores the existing anomaly detection tools by selecting those that allow continuous event flow management and those that offer heterogeneous processing of the system substrate profiles. Then, in Section~\ref{sect-case-study} a new methodology for detecting point anomalies is presented based on the graphical specification of the detection model. The methodology is illustrated through a case study in the Microsoft365 environment. The different actions that make up the model and the execution steps of this new specification will be described. Finally, in Section~\ref{sect-eval-discuss} a qualitative evaluation is proposed to show that using ASTD meets formalization, reusability, and modularity objectives that next-generation IDSs need to counter increasingly complex attacks and motley.

\section{Related work}
\label{related-work}

In the literature, some tools offer the possibility of detecting anomalies in a data set, each using a different methodology. There are those specific to anomaly detection, others more related to the processing and analysis of event logs, and others that present advanced functionalities in the statistical processing of data. 

Several industrial approaches exist to perform anomaly detection by signature or behavior. The first approach is carried out with the Snort tool. Snort~\cite{Roesch1999SnortLI} provides a low-level signature language to express and detect multi-stage Advanced Persistent Threats (APT) attacks. However, Snort is a stateless language that offers minimal event correlation capabilities. This limitation has the effect of triggering more redundant true positives and false positives. Suricata~\cite{suricata_2022} is based on the same inference mechanism as Snort, so it is very complex to make combinations to detect complex attacks. Zeek~\cite{1267552} was proposed to overcome some limitations of Snort by providing an event-driven scripting language to precisely specify and identify APT. The writing of Zeek scripts is essentially programming using functions and global variables. However, Zeek being a scripting language, is less abstract than approaches based on process algebra composition operators. Zeek functions are monolithic; that is, there is a single function for each event, and this function must address all cases of occurrence of this event, making it complex to deal with state-dependent reactions for this event.  

BeepBeep 3~\cite{bookevent} is mainly a data stream query engine. It provides processors and functions that define recurrent operations on event logs. BeepBeep 3 aims to present reusable, tested, and general toolkits that reduce the development effort of continuous event processing and express this processing in a more readable way and with a higher level of abstraction. BeepBeep 3 does not present predefined processors for anomaly detection, although such extensions exist~\cite{roudjane2018real}. BeepBeep forms more complex computations on the data by composing (or piping) processors between them, which is achieved by letting the output of one processor be the input of another. It does not present a large selection of relationships that can be established between different processors. The specification of anomaly detection is more representative and simpler by ASTDs than with BeepBeep 3. This argument means that modularity is not present with BeepBeep compared to methods based on process algebra. 

Palisade~\cite{kauffman2021palisade} is an anomaly detection framework. It is motivated by the need to remotely detect anomalies and combine a set of detectors with improving the detection system's accuracy. Palisade ensures that the different detectors can operate in parallel on the same data set thanks to its architecture composed of nodes that communicate via Redis, a distributed data streaming architecture. Palisade does not handle anomaly detection in interleaved events as it is intended for embedded systems. The detection is performed on the entity's data to which the framework is connected. Palisade does not present a graphical representation or an additional level of abstraction to develop an intrusion detection system. However, it is necessary to browse its source code to extend or reuse systems based on Palisade, making it less flexible than ASTD. 

Project-R is one of the oldest tools for statistical data processing and statistical calculations. It is a GNU project developed by the R programming language~\cite{R_1996}. It has advanced features like time series analysis, clustering, classification, etc. Thus it can be used in anomaly detection according to machine learning techniques~\cite{FLAIRS1715429,6354310}, especially at the stage of establishing the model describing the system's normal behavior. R does not offer the possibility of combining statistical processing, which causes a considerable loss of time during execution. PqR~\cite{nealspeed} improves R, whose main objective is the acceleration of calculations. PqR structure calculations as tasks by adding the possibility of parallelizing, pipelining, and merging tasks when certain conditions are met. The modularity and reusability of an IDS made with R depends on the developer. 

The management of Interleaved Event Inputs in~\cite{pao2019dealing} raises the need to separate interleaved events produced by different users or for other purposes during intrusion detection. It allows distinguishing between data elements representing different behaviors and locating where the intrusion is. Research works in~\cite{nakayama2008dynamic,sun2019strategies} indicate that the detection of anomalies in data streams and environments that dynamically change properties requires the updating of training data to preserve the accuracy of the detection system.  

The ASTD specification language, through its compiler \cASTD{}~\cite{nganyewou2020modelisation}, allows continuous data stream processing and combines the processes constituting the detection system through algebraic operators. In the following, a case study will be presented that demonstrates how to process coordination, entity processing independence, and automation of training data update can be provided by the ASTD specification of the detection system.

\section{Case Study}
\label{sect-case-study}

The case study detects unexpected events in end-user activity data streams from various Microsoft online services such as Exchange, Azure AD, and SharePoint. They are collected in real-time using a Microsoft365 API. Unexpected events occur at times of the day when the user is not usually active. Data streams are made up of events representing activities performed by various users. Among the attributes associated with an activity are: 
\begin{itemize}
    \item ID : uniquely identifies each event 
    \item CreationTime: determines the date and time in Coordinated Universal Time (UTC) that the user performed the activity. It has the following format YYYY-mm-ddTHH:MM:ssZ.
    \item  UserId : the user who performed the action 
\end{itemize}
The events are interleaved: they contain events from different users not recognized (identified) a priori (i.e., the IDS does not have access to a database of existing/registered users; it discovers them on the fly). Events are not always received in the chronological order of their realization, and some events are received very late.

Anomaly detection proceeds according to the following steps:  
\begin{itemize}
    \item We establish a model describing a user's activity during the day. This model estimates the probability density of a user's activity during the 1440 minutes of the day using the non-parametric technique kernel density estimation (KDE).
    \item A minimum threshold is set that defines the lowest probability density to classify an event as expected. 
    \item The new events are compared with the learned reference model. If the event has occurred for a minute for which the probability density is below the threshold, the event is considered to be an anomaly. 
\end{itemize}

KDE has been used in unexpected event detection in an application established in collaboration with the company Sherweb~\cite{letourneau2021statistical}. The experiments performed demonstrated that the model meets its statistical function by modeling the active hours of a user even when ignoring the exact values of the model parameters. In addition, it turns out that the reported events are abnormal in terms of user behavior and not necessarily performed by an attacker. The model's threshold is chosen considering that a significant threshold value will classify more events as abnormal, which requires more investigation by the company security analyst.

The update of the training data is done by implementing a sliding window. The events are grouped by week by assigning them a week number calculated from the DateCreation attribute, which we call henceforth a \emph{period}. A \emph{period} is defined as YYYYWW, where YYYY denotes the year and WW denotes the week's number. Two types of periods are needed: \UsedPeriods{} and \AccumulatedPeriods{}. \UsedPeriods{} are used to calculate the current KDE model, and \AccumulatedPeriods{} are the periods received after the computation of the current KDE, and that will be used to compute the next KDE.
To update the training data two conditions must be satisfied:
\begin{itemize}
    \item The accumulation of at least $n$ period
    \item Obtaining at least $k$ events in the accumulated periods
\end{itemize}
These conditions were put in place to ensure that the sample of data used for training was representative and that the profile learned by KDE was reliable.

\begin{figure}
\includegraphics[width=\textwidth,height=8cm]{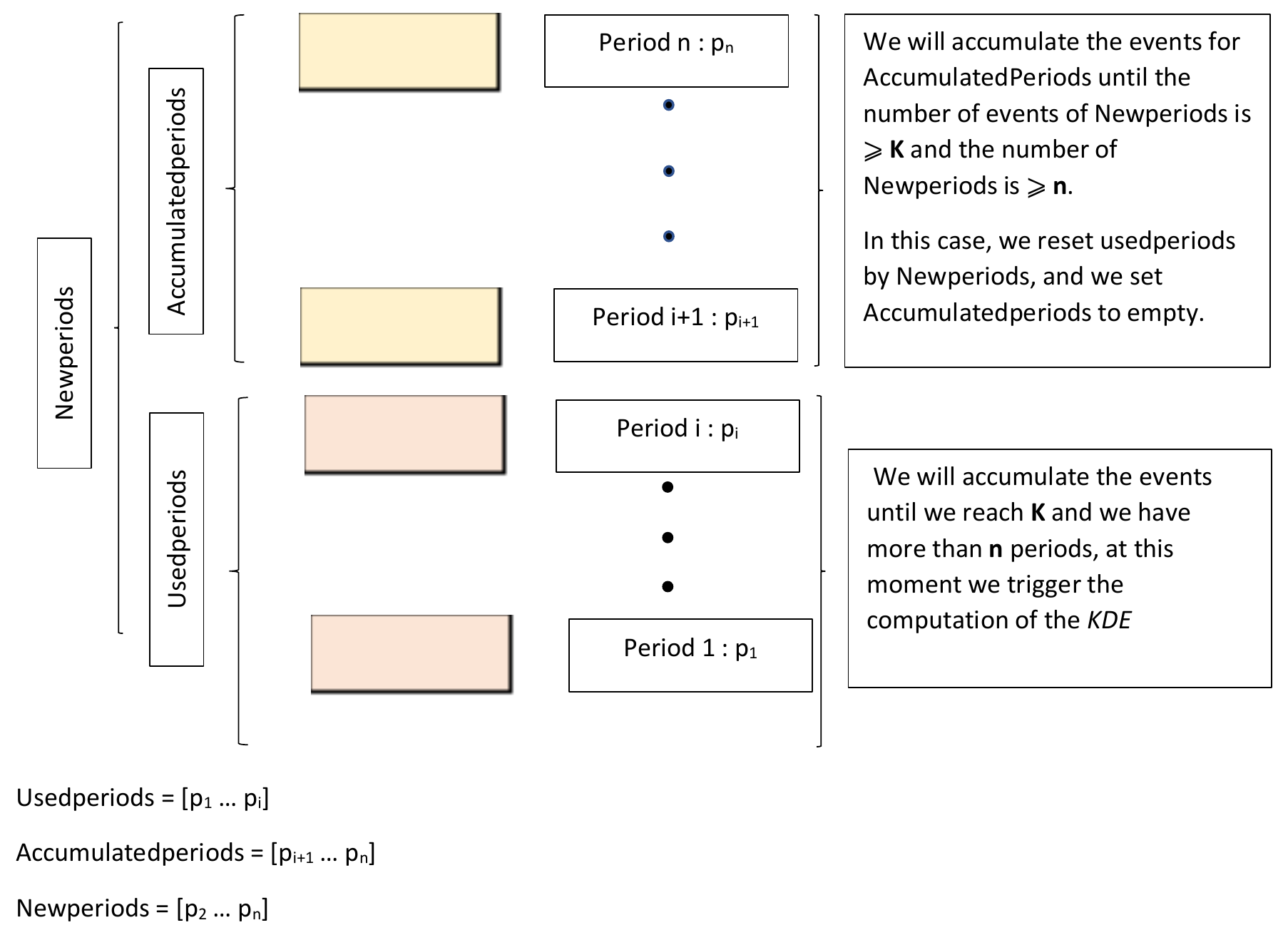}
\caption{Methodology for updating training data.} \label{fig3}
\end{figure}

Figure~\ref{fig3} represents the data renewal process.

Having accumulated at least $n$ periods in \UsedPeriods{} and obtained $k$ events associated with these periods, we launch the computation of the KDE, then we remove the first period from \UsedPeriods{} and we add the periods of \AccumulatedPeriods{} to \UsedPeriods{}. Finally, we empty the list \AccumulatedPeriods{}, and we continuously repeat this process.

In what follows, we present the graphical specification of the detection system by highlighting the process algebra operators used and their functionalities. Then we define the different actions governing the specification of the detection system and the methodology for updating the training data.


\subsection{Graphical specification of the IDS}
ASTD specifications are created using the \eASTD{} editor. The specification is built using state-transition machines, which are combined using process algebra operators, called \emph{ASTD types}. Thus, an ASTD of a given type contains an operator, attributes (i.e., state variables), and an executable code (action) which is executed every time the ASTD is executed. Each ASTD type has a specific graphical representation.

\begin{figure}[h!]
\includegraphics[width=\textwidth,height=7cm]{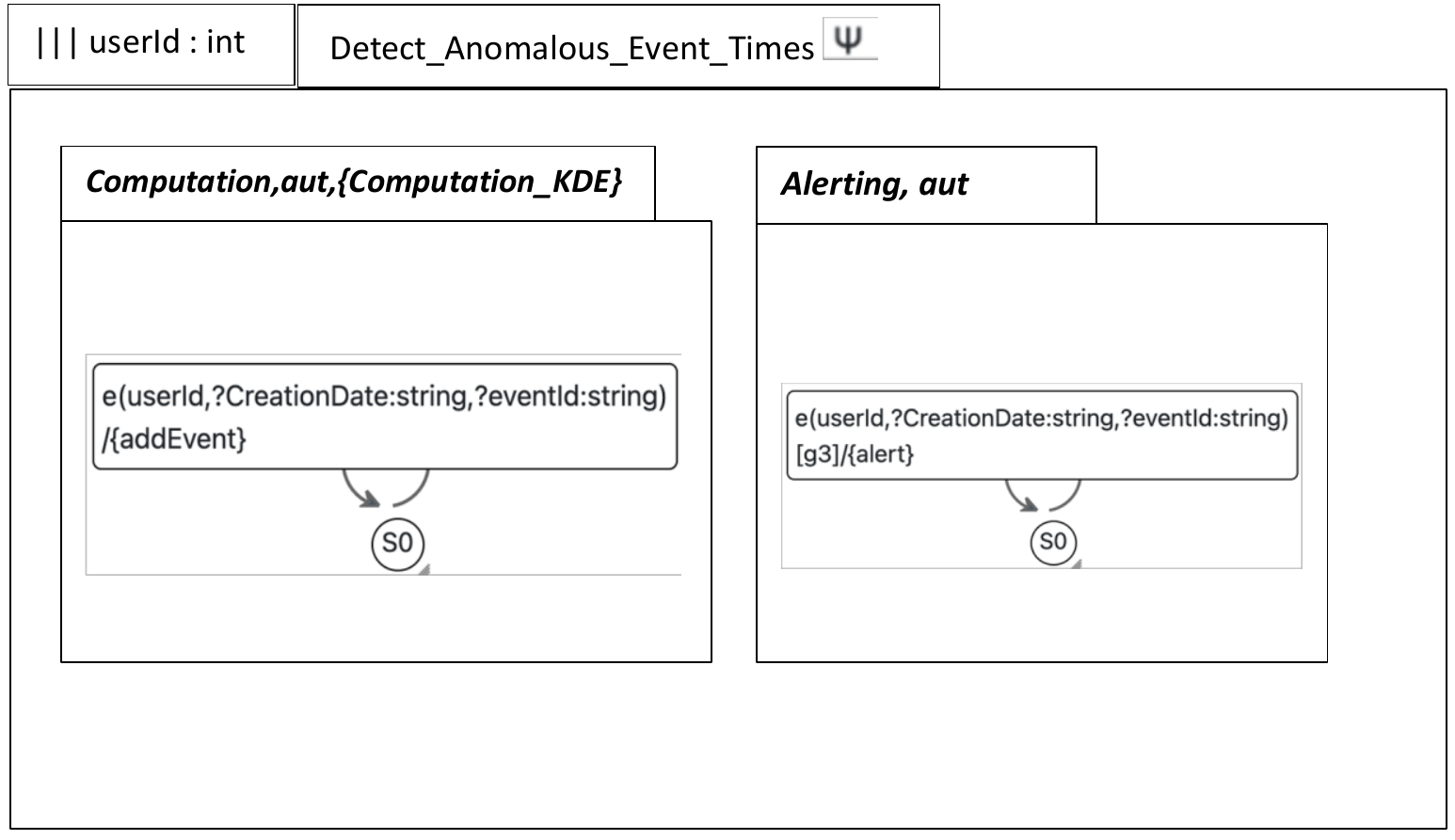}
\caption{ASTD graphical specification.} \label{fig1}
\end{figure}

Figure~\ref{fig1} provides the graphical representation of the ASTD specification of our model.  Its top-level operator is a quantified interleave, denoted by $\interleave$ in the top-left tab; it is a unary opertor, thus it applies to its sub-ASTD $Detect\_Anomalous\_Event\_Times$. It declares a quantified variable \textsf{userid} of type \textsf{int}.  ASTD was initially intended for information system (IS) modeling. The quantified interleave operator, taken from the CSP~\cite{hoare1978communicating} language, gave ASTD an advantageous property not present in other modeling languages such as UML, which consists in the possibility of representing multiple instances of the same entity in an explicit and concise way~\cite{astdarticle}. In our context, the quantified interleave operator allows one to treat each user independently by associating an instance of its sub-ASTD $Detect\_Anomalous\_Event\_Times$ to each user. Thus, each user has its own copy of this sub-ASTD, and it can store the specific information related to a user. It is important to note that the quantification variable \textsf{userid} has an unbounded domain which allows the ASTD to treat all the users without the need to recognize them before.

ASTD $Detect\_Anomalous\_Event\_Times$ is of type flow, denoted by $\downpitchfork$; it is a binary operator similar to AND-state in Statecharts.  The flow operator was added to the ASTD language in~\cite{tidjon2018extended}, because often the same event is part of several attacks, and flow allows this event to be executed on each attack specification that can execute it.  It allows for executing the same input event on both the training and detection processes.

ASTD \emph{ Detect\_Anomalous\_Event\_Times } has the following attributes:
\begin{itemize}
    \item $EventsByWeek$ : $map\langle int,vector \langle double \rangle\rangle$ ; it contains the period as a key, and a list of event minutes.
    \item $n$ : $int$ ; the minimum number of periods to accumulate to launch the calculation of the KDE  
    \item $k$ : $int$ ; defines the number of events that a user should have in $n$ periods, in order to compute the KDE and build his profile.
    \item $threshold$ : $double$; defines the lowest probability to classify an event as expected
    \item \UsedPeriods{} : $vector \langle int \rangle$ ; it contains the indices of the periods in EventsByWeek which will provide the calculation of the KDE after having accumulated a minimum of $K$ events for these periods. 
    \item \AccumulatedPeriods{} : $vector \langle int \rangle$ ; it is used to renew the data used for the calculation KDE. 
    \item $\emph{startKDE}$ : $bool$ ; is used to launch the KDE calculation when it is true. 
    \item $\emph{UserKDE}$ : $vector \langle double \rangle$ ; it contains the current KDE calculated. 
    \item $\emph{Alerts}$ : $vector \langle string \rangle $ ; it contains the ID of the suspicious events.  
\end{itemize}

\emph{Detect\_Anomalous\_Event\_Times } contains two sub-ASTDs: \emph{Computation} and \emph{Alerting}, which in turn have access to the previous attributes. The event $\emph{e}$ is executed by each sub-ASTD which can execute it.

The ASTD \emph{Computation} is of type Automaton. It has as an action  \emph{KDE\_Com\-putation} which takes as parameters the following variables and attributes: \emph{userId, EventsByWeek, userkde, \UsedPeriods{}, \AccumulatedPeriods{}, startkde}; it is responsible for the KDE computation after checking the value of \emph{startKDE}. ASTD \emph{Computation} is a state machine that contains a single state with a loop transition labeled with event $\emph{e}$ and it has an action \emph{addEvent(userId, CreationTime, EventsByWeek, UsedPeriods, AccumulatedPeriods, startkde, k, n)}, which adds the events received to the map $EventsByWeek$ and manages the periods. The execution of actions occurs in a bottom-up way, which means that transition actions are executed first, followed by ASTD actions. Thus, action  \emph{addEvent} is executed before action \emph{KDE\_Computation}. ASTD \emph{Computation} manages the attribute $EventsByWeek$ and the computation of the KDE profile when it is possible.

The ASTD \emph{Alerting} is also a state machine. It contains only one state with a loop transition also labeled with \emph{e}, and it has an action \emph{alert(userkde, userId, CreationDate, ID, alerts, threshold)}, which is in charge of checking if the probability of occurrence of the received event is lower than the threshold. In that case, the event is reported by adding its $ID$ to the vector of alerts. The transition is guarded with condition $\emph{g3} = \emph{userkde. size()!=0}$, which ensures that the \emph{userkde} is not empty.

\subsection{Action Definitions}

First of all we define the three main actions $(\emph{addEvent}, \emph{Computation\_KDE}, \emph{alert})$, then we introduce some methods responsible for partial calculations.

\begin{algorithm}[h]
\caption{addEvent}\label{addEvent}
\hspace*{\algorithmicindent} \textbf{Input:}$userId$, $CreationDate$, $EventsByWeek$, $\UsedPeriods{}$, $\AccumulatedPeriods{}$, $startKDE$, $n$, $k$   \\
\hspace*{\algorithmicindent} \textbf{Output:} $EventsByWeek$, $\UsedPeriods{}$, $\AccumulatedPeriods{}$, $startKDE$ updated
\begin{algorithmic}[1]
\State $period \gets Compute\_period(CreationDate)$
\State $value \gets Compute\_minute(CreationDate)$
\State $EventsByWeek[period].append(value)$
\If{$\UsedPeriods{}.size()!=0$}
\State $last\_used\_period \gets \UsedPeriods{}[\UsedPeriods{}.size() - 1]$
\EndIf
\If{$calculNbrEvents(EventsByWeek, \UsedPeriods{})\leq k \; \textbf{or} \;\UsedPeriods{}.size() \leq n \; \textbf{or} \; diffnext(last\_used\_period, period) > 0$}
\If{$period \; \textbf{not in} \; \UsedPeriods{}$}
\State $insert(\UsedPeriods{},period)$
\EndIf
\ElsIf{$period \; \textbf{not in} \; \UsedPeriods{}$}
\If{$\AccumulatedPeriods{}.size() == 0$}
\State $startKDE \gets true$
\EndIf
\If{$period\; \textbf{not in} \;\AccumulatedPeriods{}$}
\State $insert(\AccumulatedPeriods{},period)$
\EndIf
\EndIf
\State $\newperiods{} \gets \UsedPeriods{}[2:] + \AccumulatedPeriods{}$
\If{$calculNbrEvents ( EventsByWeek, \newperiods{})\geq k \; \textbf{and} \; calculNbrEvents(EventsByWeek, \AccumulatedPeriods{}) \geq 2  \; \textbf{and} \; \UsedPeriods{}.size() \geq n$}
\State $EventsByWeek.erase(\UsedPeriods{}[1])$
\State $\UsedPeriods{} \gets \newperiods{}$
\State $\AccumulatedPeriods{} \gets []$
\State $\newperiods{} \gets []$
\EndIf
\end{algorithmic}
\end{algorithm}

Action \emph{$addEvent$} (see Algorithm~\ref{addEvent}) updates the training data structure ($\emph{EventsByWeek}$) and triggers the KDE computation.
For each event received, we calculate the minute of the day and the period in which it occurred from the $\emph{CreationDate}$ by the $\emph{Compute\_minute}$ and $\emph{Compute\_period}$ methods, respectively. The minute obtained is then added to the $\emph{EventsByWeek}$ map according to its period. The condition of line 6 allows to build the list \UsedPeriods{} and to ensure the continuity of the order between \UsedPeriods{} and \AccumulatedPeriods{} by verifying that the inserted period is less than the last period of \UsedPeriods{}. If this condition (in line 6) is not satisfied, it means that the computation of the KDE from the data associated with \UsedPeriods{} is possible. To ensure that we have received enough or all events from the last \UsedPeriods{} period, we check-in line 10 that we have not yet received an event from a brand new period that does not exist in \UsedPeriods{}. The condition in line 12 ensures that the \AccumulatedPeriods{} list is built until the conditions for updating the training data are satisfied.

Then we create the \newperiods{} list by taking the \UsedPeriods{} list deprived of its first period and the periods of \AccumulatedPeriods{}. The condition in line 15 checks if \newperiods{} can be the new \UsedPeriods{} that will be used for the computation of the new profile and that there are at least two events associated with the \AccumulatedPeriods{}; this is to ensure that the first period of \UsedPeriods{} is not deleted before being included in the KDE calculation because the first event of the \AccumulatedPeriods{} is responsible for starting the KDE  computation.
 
\begin{algorithm}[h!]
\caption{Computation\_KDE}\label{calculKDE}
\hspace*{\algorithmicindent} \textbf{Input:} $EventsByWeek$, $\UsedPeriods{}$, $\AccumulatedPeriods{}$, $startKDE$, $userKDE$ \\
\hspace*{\algorithmicindent} \textbf{Output:} $userKDE$ updated
\begin{algorithmic}[1]
\If {$startKDE$}
\State $userKDE$.clear() \texttt{\Comment{reset $userKDE$}}
\For{key in $EventsByWeek$.keys()}
\If{key not in $\UsedPeriods{}$ \textbf{and} key not in $\AccumulatedPeriods{}$}
\State $EventsByWeek$.erase(key)
\EndIf
\EndFor
\State $fusion(EventsByWeek,\, \UsedPeriods{},\, fusiondata)$
\State $userKDE \gets computation of the KDE$
\State $startKDE \gets false$
\EndIf
\end{algorithmic}
\end{algorithm}

Action $\emph{Computation\_KDE}$ (See algorithm~\ref{calculKDE}) computes the KDE after verifying the value of $startKDE$. In this case, it resets the $userKDE$, cleans up the map \emph{EventsByWeek} by deleting the periods not existing in $\UsedPeriods{}$ and $\AccumulatedPeriods{}$, merges the data in $\emph{EventByWeek}$ from the \UsedPeriods{} into a single list and starts the KDE computation. 

\begin{algorithm}[h!]
\caption{alert}\label{alert}
\hspace*{\algorithmicindent} \textbf{Input:} $userKDE$, $ID$, $CreationDate$, $alerts$, $threshold$ \\
\hspace*{\algorithmicindent} \textbf{Output:} $alerts$ updated
\begin{algorithmic}[1]
\State $value \gets Compute\_minute(CreationDate)$
\If{$userKDE[value] \leq threshold$}
\State add $ID$ to $alerts$
\EndIf
\end{algorithmic}
\end{algorithm}

Action $\emph{Alert}$ (see Algorithm~\ref{alert}) compares the probability of occurrence of the event and the threshold. It computes the minute of occurrence of the received event. It retrieves the probability of occurrence of events at this minute using $userKDE$, compares the probability to a threshold. If the probability is less than the threshold, $ID$ is added to the alerts list.

The $numweek$ method  receives as input the day, month and year. It returns the number of the week associated with this date.        
The $calculNbrEvents$ method receives as input a map of event data and a list of periods. It returns the number of values (events) for those periods in the map. 
The $\mathit{diffnext}\langle period1, period2 \rangle$ method calculates the difference between two periods, assuming that period2 $>$ period1.

The $\emph{insert}$ method (See Algorithm~\ref{insert}) inserts the periods in the lists $\UsedPeriods{}$ and $\AccumulatedPeriods{}$. The insertion of the periods in the two lists is done while keeping an ascending order. This order in the lists is created to ensure the order of the events and identify the events received late: It sometimes happens that there are events that are received after one month of their occurrence. Therefore, it would be relevant to delete them when calculating the current KDE profile. These events can be determined because they belong to a period very far from the first period of the current period list (a difference of more than three periods between the old period and the first period of the list). If we receive an event corresponding to an old period, this period will not be inserted into the list. 

\begin{algorithm}[h]
\caption{insert}\label{insert}
\hspace*{\algorithmicindent} \textbf{Input:} $vec$, $period$ \\
\hspace*{\algorithmicindent} \textbf{Output:} $vec$ updated
\begin{algorithmic}[1]
\State $it \gets \textit{upper\_bound}(vec.begin(), vec.end(), period)$\texttt{\Comment{$\textit{upper\_bound}$ return an iterator pointing to the first period in the range  [vec.begin(),vec.end())  which compares greater than period}}
\If{$it == vec.begin()\; \textbf{and} \; it != vec.end()$} \texttt{\Comment{if vec is not empty and period should be inserted at the beginning of vec}}
\State $diff \gets diffnext(\textit{value at it}, period)$
\If{$diff \leq 3$} 
\State insert $period$ at the position pointed by $it$
\EndIf
\Else
\State insert $period$ at the position pointed by $it$
\EndIf
\end{algorithmic}
\end{algorithm}

\subsection{IDS code generation}
The generation of the IDS source code is done by compiling the ASTD specification by the \cASTD{} compiler~\cite{tidjon2020formal}. The latter produces code in C++ programming language from an ASTD specification in JSON. The compilation takes place in the following four steps :
\begin{itemize}
    \item Parsing the ASTD specification in JSON and producing an ASTD object model by the ASTD Parser.
    \item Translation from ASTD to an intermediate model (IM) using the ASTD Compiler.
    \item Translation from IM to a programming language like C++ using the IM Translator.
    \item Code optimization by removing redundant calculations
\end{itemize}

\begin{figure}[h!]
\includegraphics[width=\textwidth, height = 9cm]{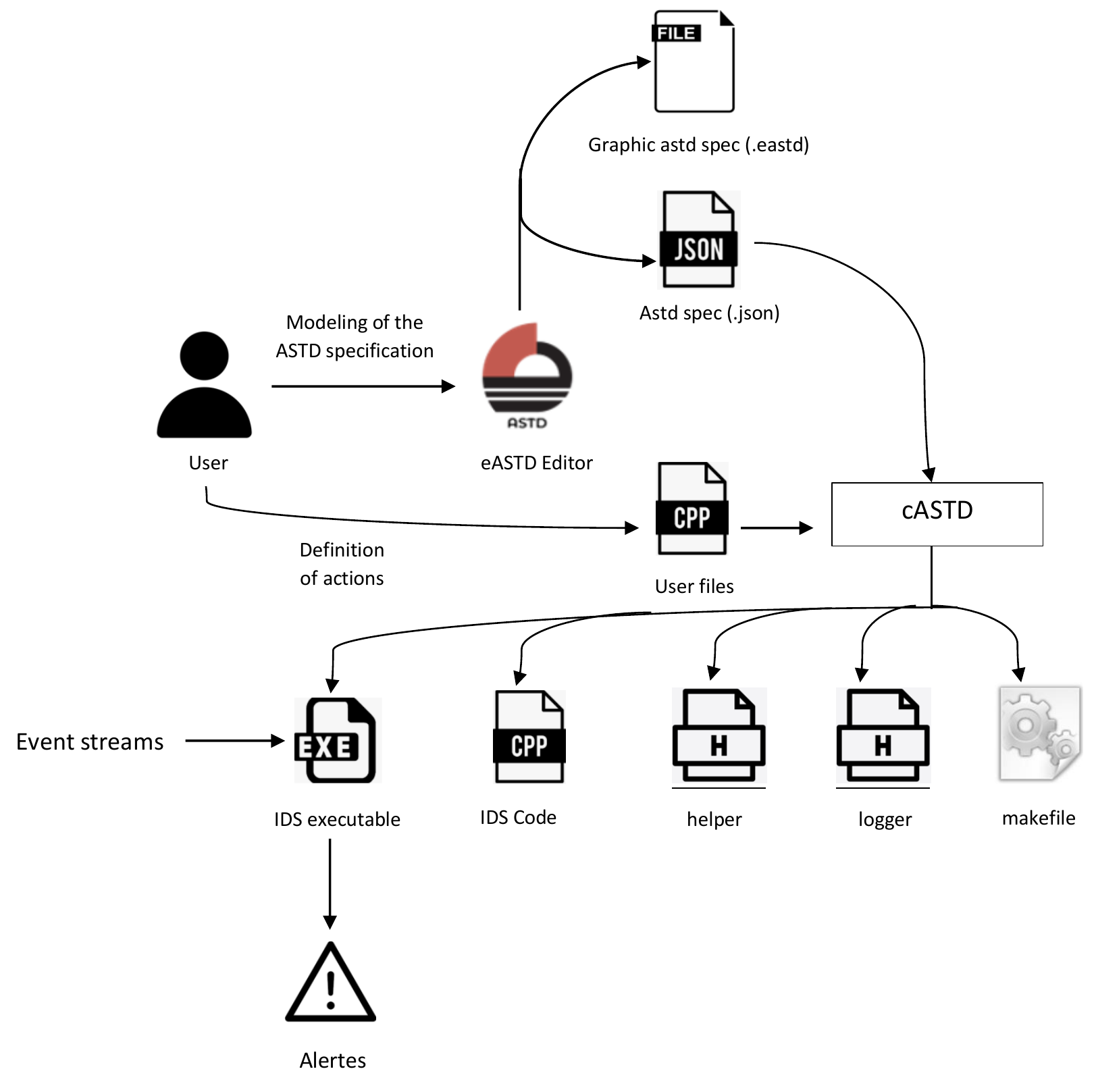}
\caption{IDS code generation.} \label{fig2}
\end{figure}

The specification is first modeled using the \eASTD{} editor, which generates the specification in JSON. This specification and the code defining the set of operations required for the training and detection processes are passed as input to the \cASTD{}. It generates as output the source code in the C++ programming language and the associated program (monitor) that will be executed on the data streams (see Fig~\ref{fig2}). The source code is composed of the helper file, which calls the constructors associated with given string types; the logger file, which allows debugging of the generated program; the IDS source code file, which contains the translated code of the ASTD specifications; and the makefile for linking and compilation. This makefile calls the native compiler corresponding to the C++ language and is automatically executed by \cASTD{} to produce the IDS executable.

\subsection{Example of specification execution}

To clarify the period management methodology, we proceed with an explicit example. It is assumed that a user's events are received with the following sequence of periods: [202225, 202225, 202225, 202221, 202227, 202227, 202227, 202228, 202228, 202228,202228 , 202229, 202229, 202226]. We take $k = 10$, $n = 3$ and $threshold = 0.001$.

When the first three events are received, the lists of periods are as follows: $\UsedPeriods{} = [202225 \;(3\ events)]$, $\AccumulatedPeriods{} = []$ and $\newperiods{} = []$, according to the condition in line 6 of the $\emph{addEvent}$ method

We receive the event of period 202221, which is supposed to be inserted in the first position in \UsedPeriods{}, in order to keep the list in ascending order. However when we compute the difference between the first period of \UsedPeriods{} 202224 and 202221, we obtain 4 which is greater than 3 so the period 202221 will not be inserted in \UsedPeriods{} according to the method insert (Algorithm~\ref{insert}). The lists of periods remains unchanged: $\UsedPeriods{} = [202225\;(3 events)]$, $\AccumulatedPeriods{} = []$ and $\newperiods{} = []$.

The following seven events [202227, 202227, 202227, 202228, 202228, 202228, 202228] will be inserted in \UsedPeriods{} according to the method $\emph{addEvent}$ (Algorithm~\ref{addEvent}). The lists of periods will have the following content: \UsedPeriods{} = [202225 \;(3 events), 202227 \;(3 events), 202228 \;(4 events)], \AccumulatedPeriods{} = [] and \newperiods{} = [202227\;(3 events), 202228\;(3 events)].

We receive the event of the period 202229, $\UsedPeriods{}$ contains more than $n$ periods, the number of events associated to it is equal to $k$, the last period of $\UsedPeriods{}$ 202228 is less than 202229, which means that period 202228 is finished, and $\AccumulatedPeriods{}$ is empty. So all the conditions are satisfied to launch the computation of the KDE : $\emph{startKDE}$ receives true. According to the condition in line 11 of $\emph{addEvent}$, we add 202229 to $\AccumulatedPeriods{}$. The period lists are as follows: \UsedPeriods{} = [202225\; (3 events), 202227\; (3 events), 202228 \;(4 events)], $\AccumulatedPeriods{} = [202229\; (1 event)]$ and \newperiods{} = [202227 \;(3 events), 202228 \;(3 events), 202229 \;(1 event)]. Then the $\emph{Computation\_KDE}$ method is executed to compute the user profile that will be stored in $\emph{UserKDE}$.

We receive another event of the period 202229. The content of the periods is as follows:
\UsedPeriods{} = [202225 \;(3 events), 202227\; (3 events), 202228 \;(4 events)], \AccumulatedPeriods{} = [202229\; (2 event)] and \newperiods{} = [202227\; (3 events), 202228\; (3 events), 202229 \;(2 event)]. The condition in the line 15 of $\emph{addEvent}$ is satisfied, we renew the list of periods to have the following:
\UsedPeriods{} = [202227\; (3 events), 202228\; (3 events), 202229 \;(2 event)], \AccumulatedPeriods{} = [] and $\newperiods{} = []$. The condition $\emph{g3}$ is satisfied which allows to execute the $\emph{Alert}$ method which classifies the event as being normal or abnormal.

Finally we receive the event of the period 202226, which, as the condition of the line 6 of $\emph{addEvent}$, is inserted in $\UsedPeriods{}$, to obtain the following lists: 
\UsedPeriods{} = [202226\; (1 event), 202227\; (3 events), 202228 \;(3 events), 202229\; (2 events), 202229 \;(2 events)], 
\AccumulatedPeriods{} = [] and \newperiods{} = [202227\; (3 events), 202228 \;(3 events), 202229 \;(2 events), 202229 \;(2 events)]. This event also passes through the detection process as  $\emph{g3}$ is satisfied.

\section{Evaluation and Discussion}
\label{sect-eval-discuss}

The flow operator coordinated the two sub-ASTDs of the training and detection processes. A received event is added to the training data structure and evaluated against the learned model if already computed. This means that the training data is fed simultaneously as the detection is maintained. The fact that the two sub-ASTDs share the attributes inherited from the parent ASTD reinforces this coordination, as the model computed by the first sub-ASTD is used to perform the detection at the second sub-ASTD.

The quantified interleave operator creates the sub-ASTD for each entity processed, which means that there are as many independent IDS as there are entities. The advantage of the processing independence of the different users in the developed case study is that the update of the training data is done depending on the user's activity, not blindly at the same time for all entities. Each entity has its own attributes, and in the ASTD language, an attribute can be initialized by the value returned by a method. The initialization can also be dependent on the properties of the entity. 

The automation of the training data update has been implemented according to if-else instructions; this is possible thanks to continuous data processing. The reference profile is renewed when a new training data set meets defined conditions. This was achieved in the case study by the executable code $\emph{Computation\_KDE}$, responsible for recalculating the KDE profile according to the value of the boolean $startKDE$, which is executed directly after the execution of the action $\emph{addEvent}$  that checks the value of $startKDE$.

The ASTD language formalized the scheduling and coordination of the various IDS processes, which reduces the mental load and the effort required for the development. The modeling of the specification in a graphical representation facilitates modifying or extending the specification. The \cASTD{} tool with its IDS source code generation methodology ensures the modularity and reusability of the IDS specification. Modularity results from the separation of the reading of the data flows, the operations composing the system, and the coordination of the different processes. Reusability is due to the fact that the IDS code generated by \cASTD{} can be compiled and executed in any environment.

The IDS specification was executed on a dataset containing 3,827,551 events of 10 distinct users collected over 10 weeks from one of our industrial collaborators operating a Microsoft365 SAS, using an Intel Core i7 processor machine with frequency 3 GHz x 8 and 32 GB of RAM. The execution took 1h:34min:13s (wall clock time) and 138MB of RAM, which represents an average execution time of 1.476ms by event.
This Microsoft365 site collects around 40 million events per week, so our generated code can clearly cope with this workload.
In~\cite{tidjon2020formal} a comparison was made between \cASTD{} and other event processing tools (BeepBeep v3, MonPoly, iASTD); it turned out that \cASTD{} is the fastest among them.

\section{Conclusion}

A study on the use of ASTDs in the context of intrusion detection systems is presented in this paper. This study demonstrates the simplicity of allowing data updates without interrupting the detection process and the creation of modularity at the level of each user that can be treated independently. The ASTD language made it possible to coordinate the anomaly detection model's different processes by using the algebra operators. Evaluation conducted in this paper shows that ASTD specification language can be used to develop anomaly-based detection systems. The \cASTD{}{} tool that has been used to compile the specifications has shown to be efficient in terms of execution time.

The anomaly detection application presented in the case study is of point type detection. Future works will be carried out to develop other types of anomaly detection that are contextual or collective. A new evaluation will present the advantages of the ASTD specification language in terms of flexibility to apply other estimation calculation methods. The hope founded by this formalization work can allow greater resilience of detection systems in the face of new modern threats. Moreover, the work carried out in this study shows that this could be a probable solution to the current detection problems.

%
%
%
\bibliographystyle{splncs04}
\bibliography{bibliography.bib}

\end{document}